\documentclass{article}
\usepackage{ijcai18}

\usepackage{times}
\usepackage{xcolor}
\usepackage{soul}
\usepackage[utf8]{inputenc}
\usepackage[small]{caption}

\usepackage{amssymb}
\usepackage{amsmath}
\DeclareMathOperator{\Tr}{Tr}
\usepackage{graphicx}
\usepackage{multirow}

\title{User-Sensitive Recommendation Ensemble with Clustered Multi-Task Learning}

\author{
Menghan Wang$^1$, 
Xiaolin Zheng$^1$, 
Kun Zhang$^2$, \\ 
$^1$ Zhejiang University \\
$^2$ Carnegie Mellon University\\
mewang@andrew.cmu.edu,
xlzheng@zju.edu.cn,
kunz1@cmu.edu
}

\begin{document}

\maketitle

\begin{abstract}
This paper considers recommendation algorithm ensembles in a user-sensitive manner. Recently researchers have proposed various effective recommendation algorithms, which utilized different aspects of the data and different techniques. However, the ``user skewed prediction'' problem may exist for almost all recommendation algorithms -- algorithms with best average predictive accuracy may cover up that the algorithms may perform poorly for some part of users, which will lead to biased services in real scenarios. In this paper, we propose a user-sensitive ensemble method named ``UREC'' to address this issue. We first cluster users based on the recommendation predictions, then we use multi-task learning to learn the user-sensitive ensemble function for the users. In addition, to alleviate the negative effects of new user problem to clustering users, we propose an approximate approach based on a spectral relaxation. Experiments on real-world datasets demonstrate the superiority of our methods.
\end{abstract}

\section{Introduction}
In recent years recommender systems have become increasingly popular, and various effective recommendation algorithms have been proposed. 
Currently one trend of recommendation research is to incorporate collaborative filtering with side information (e.g, social information \cite{wang2017collaborative} and item reviews \cite{mcauley2013hidden}) and other promising techniques (e.g., deep learning \cite{karatzoglou2017deep} and transfer learning \cite{weiss2016survey}).

However, few studies have focused on the ``skewed prediction'' problem, where the model with the best average predictive accuracy will leave meaningful subsets of users/items modeled significantly worse than other subsets \cite{beutel2017beyond}. The ``skewed prediction'' problem may become a common drawback of the existing recommendation algorithms as they use average based metrics for evaluation. The globally optimal model is typically not the best model for all the users. We focus on the user skewness and show an illustrative experimental result on the public dataset \emph{MovieLens-100K} in Figure 1. Traditionally we compare the average MSE of two algorithms, and from the left table we can easily tell the MF model is better. But if we analyze the MSE on user level, we can see that only 53.2\% of users suit MF more than KNN. Directly deploying MF for real use will provide biased services to users. Besides the performance measurement, one important reason is user heterogeneity. Specifically, there are many types of users and an algorithm with predefined structure and techniques probably can not capture the preference of all types of users well, which accounts for the user skewed predictions. This problem may also exist among recent advanced recommendation algorithms. For example, social recommendation utilizes user's social information to capture user preference, and it will underperform for the users who have few social friends. Tag-based recommendation also has similar problems as not all the users like to annotate tags. 

\begin{figure}[t]
\centering
\includegraphics[width=8.0cm]{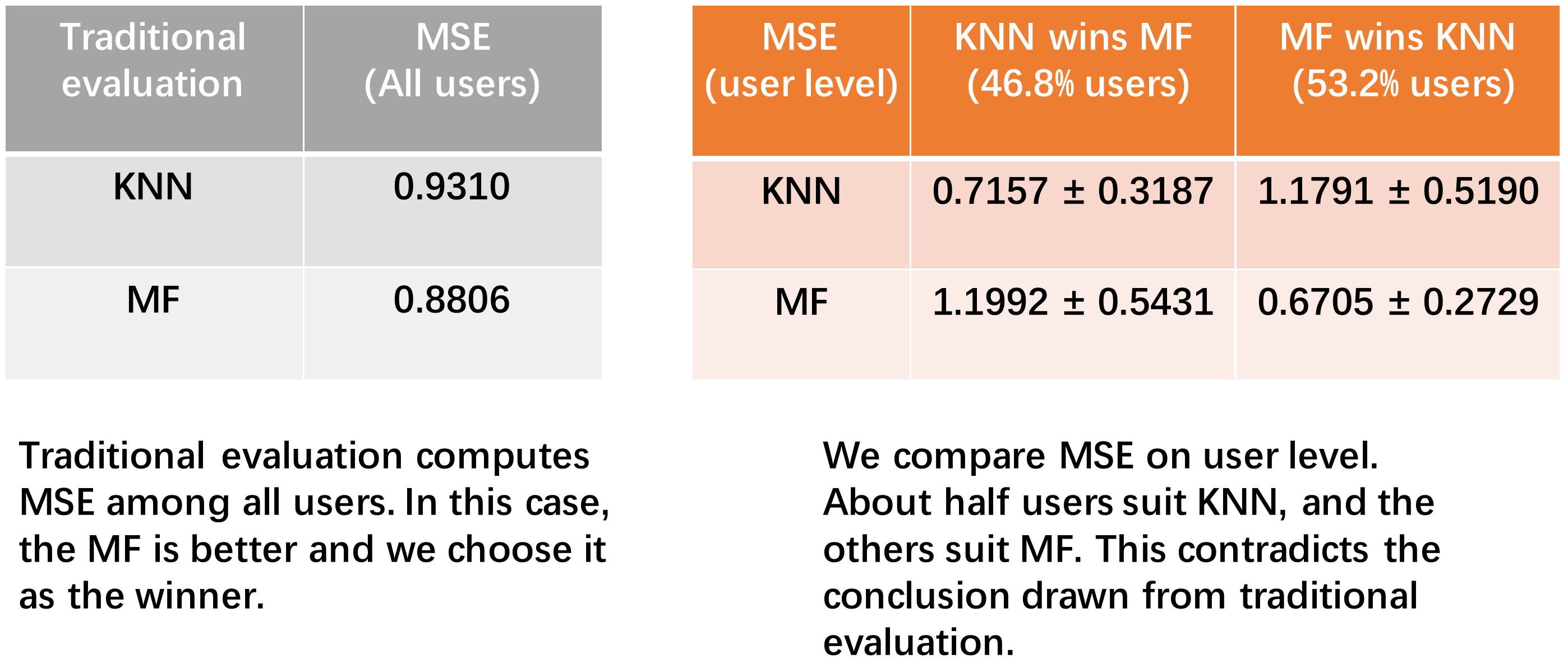}
\caption{A comparison of mean square error (MSE) on two classical recommendation algorithms K-nearest neighbors (KNN) and Matrix Factorization (MF) on the \emph{MovieLens-100K} dataset. }
\label{example}
\end{figure}

This paper considers recommendation ensembles to address this problem. Different algorithms may suit different types of users and a natural way to improve the recommendation performance is to combine them properly. Traditional ensemble methods (e.g., bagging, boosting, and stacking) often learn a weight for each base algorithm and apply it to all users, which, however, do not consider the user heterogeneity phenomenon and thus are not the optimal choice in the recommendation field. 
We assume that homogeneous users should share similar ensemble strategies as algorithms are expected to perform stably on homogeneous users. Then a natural idea is to divide users into several homogeneous groups by analyzing base algorithms and learn ensemble weights within each group. The intuition is clear and sensible: if recommendation algorithms have similar performances on some users, these users are more likely to be homogeneous and they should share similar weights during ensemble. This idea is very close to clustered multi-task learning \cite{zhou2011clustered} if we treat ensemble for each user as a single task. Users follow a clustered structure and users in the same group share parameters during task learning. Compared to a more personalized approach that learns ensemble for users individually, our method tends to be more reliable and can alleviate the data sparsity problem.  We call this ensemble strategy is user-sensitive.

In this paper we propose a user-sensitive recommendation ensemble approach, named ``UREC", to address the ``user skewed prediction'' problem. The main contributions of our work are listed as follows: (1) We first cluster users into homogeneous groups, and then use multi-task learning to learn ensemble function strategies for the users. To our best knowledge, it is the first work to consider user heterogeneity in recommendation ensemble. (2) To alleviate the new user problem that will interfere with the user clustering, we propose an approximate approach based on a spectral relaxation of regularization. (3) We conduct extensive experiments on real-world datasets to verify the efficacy of our method. The experimental results demonstrate that UREC outperforms other baseline models.

\section{Related Work}

\subsection{Multi-task Learning}
Multi-task learning (MTL) \cite{ando2005framework} is a machine learning method where multiple tasks are jointly learnt such that each of them benefits from each other. Several researchers have applied MTL to recommendation with different assumption on how to define a task and what to share among tasks. \cite{ning2010multi} proposed a multi-task model for recommendation with Support Vector Regression. But they focused on the task on the individual level and only used rating information. \cite{wang2013online} utilized MTL to online collaborative filtering where the weight vectors of multiple tasks are updated in an online manner. These works assumed that all the tasks are related. However, we assume a more sophisticated group structure among users where users only share relatedness within the same group. 

\subsection{Recommendation Ensemble}
Ensemble-based algorithms have been well studied to improve the prediction performance \cite{polikar2006ensemble}, and are widely adopted in recommendation competitions, such as the Netflix Prize contest \cite{sill2009feature,koren2009bellkor} and KDD Cups \cite{DBLP:journals/jmlr/McKenzieFCLTWCLLYLWNSKTCCCCWWLLL12}. Typically, an ensemble method combines the results of different algorithms to obtain a final prediction. The most basic strategy is to acquire the final prediction based on the mean over all the prediction results or the majority votes. Some popular ensemble methods are linear regression, restricted boltzmann machines (RBM), and gradient boosted decision trees (GBDT)  \cite{polikar2006ensemble}. However, they assume users are homogeneous and use the same ensemble strategy to all the users. It is then desirable to develop a user-sensitive ensemble method to capture and make use of user heterogeneity, as we shall do next.
\subsection{Hybrid Recommendation Algorithms}
Another related field is hybrid recommendation. Different from recommendation ensembles that combines the results of different algorithms, hybrid recommendation aims to build a model with multiple recommendation techniques to achieve a higher performance. Hybrid recommendation models have shown competitive results.
The most common hybrid recommendation is to combine collaborative filtering with other techniques like content based model \cite{basilico2004unifying}, clustering \cite{hu2014clubcf}, and Bayesian model \cite{beutel2014cobafi}. 
One potential drawback of hybrid recommendation is the model structure and inference rules will become sophisticated when more techniques come into consideration. However, our method combines multiple techniques in an ensemble approach that can avoid this problem.

\section{User Sensitive Recommendation Ensemble}

\begin{table}[t]
\small
\label{table_symbol}
\centering
\begin{tabular}{lccc}
\hline
 Symbol &  Description  \\
\hline
\(Y_{ij}\) & the rating user \(i\) gives to item \(j\) \\
\(X^{k}\)  & the full predicted matrix by algorithm $k$\\
\(W\) & the ensemble weight matrix that is \([w_{1},..., w_{N}] \)\\
\(A^{k}\) & the \(k\)th base algorithm, \(k \in [1,2,...,K] \)\\
\(Z\) & user type (homogeneous group) \\
\(\bar{w}_{z}\) & the average ensemble weights of group $z$\\
\(m^{k}_{i}\) & metric performance of algorithm $k$ on user $i$ \\
\(d^{k}_{iq}\) & distance between users $i$ and $q$ based on algorithm $k$  \\
\(\alpha,\beta\)  & the regularization parameters \\

\hline
\end{tabular}
\caption{Notation}
\end{table}

\subsection{Problem Description}
Suppose there are \(N\) users and \(M\) items, and \(K\) recommendation algorithms. Let \(Y \in \mathbb{R}^{N \times M}\) be the user-item rating matrix where \(Y_{ij}\) represents the preference user \(i\) towards item \(j\). Note that in most cases the \(Y\) is very sparse -- most values in \(Y\) are missing. For every recommendation algorithm \(k\), there is a prediction matrix \(X^{k} \in \mathbb{R}^{N \times M}\)  that stores all the predicted preferences of \(N\) users towards \(M\) items. Ensemble learning is to find a model that can better predict \(Y\) based on \(\{X^{k}\}\) without revising the inner design of the \(K\) recommendation algorithms. Existing studies ignored the difference between users and treated them with same weights. In this paper, we use an adaptive ensemble model for all users, and our goal is to improve the overall prediction accuracy of the ensemble model by developing user-sensitive model parameters for users. We give the notation in Table 1.

\subsection{Proposed Model}
We introduce the formalization of UREC and discuss intuitions in detail with probabilistic graphical models in Figure 2. Figure 2(a) shows relationships among user hidden types $Z$, user features $F$, recommendation algorithms $A$, and user-item rating matrix \(Y\).   User features \(F\) are determined by the hidden user type Z, and predictions \(X^{k}\) are generated by algorithms \(A^{k}\) and user features \(F^{k}\). Note that in reality \(F^{k}\) is often a very small subset of \(F\). Most current algorithms only use users' historical ratings as input. 
When (some or all of) the features  are not observable, they are integrated out in the graphical representation given in (a); as a consequence, $X^k$ and $Y$ will be conditionally dependent (even given $Z$), as shown by the dashed line in Figure 2(b). 
Our purpose is to leverage all $X^k$'s for a better prediction of $Y$, as supported by the dependence between al $X^k$'s and $Y$. 
This simplified graphical representation indicates that  \(P(Y|X^{K}) = \sum_{z=1}^{Z}  P(Y|X^{K},z) P(z)\), where \(X^{K} = \{X^{k}\}\). Traditional ensemble methods assume users are homogeneous, that is \(P(Y|X^{K}) = P(Y|X^{K},Z )\), and use \(P(Y|X^{K})\) for all users. In the case of heterogeneous users, \(P(Y|X^{K},z)\) may be different for different \(z\). So our UREC involves two steps: the first step is to divide users into \(Z\) homogeneous groups and the second step is to learn automated ensemble strategies within each group. Alternatively we can use soft clustering like Gaussian mixture models so that we can model user homogeneity more continuously. However, the high dimensions of \(X^{K}\) and \(Y\) will make the probabilistic clustering hard to deal with. So here we apply a k-medoids method based on the performance of base algorithms, which is hard clustering and detailed in the next subsection. In the second step we choose linear mixtures with automated determined coefficients as the ensemble model, which can be further extended to other models like trees and neural networks.


\begin{figure}[t]
\centering
\includegraphics[width=8.0cm]{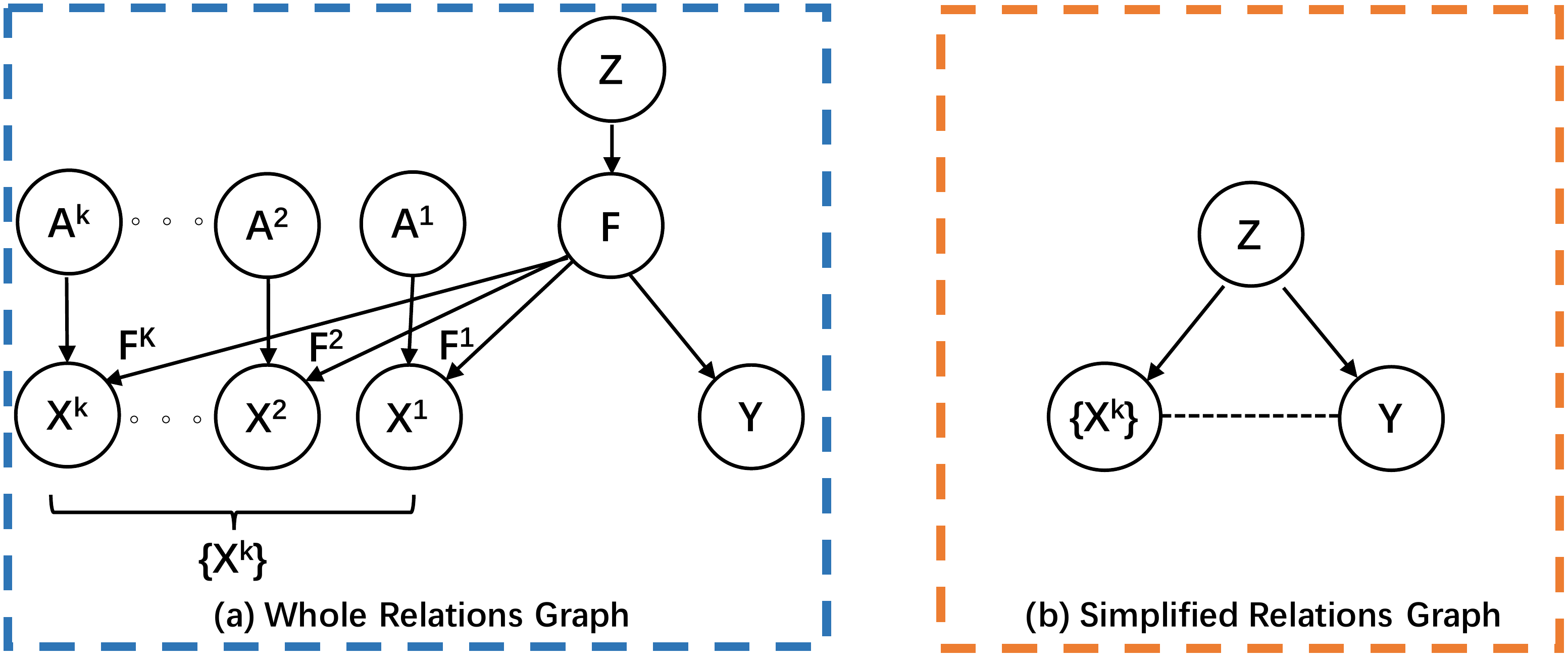}
\caption{Relations graphs of recommendation. }
\label{model}
\end{figure}

\subsection{Weighted User Grouping}
In this section we introduce the weighted user grouping method, where the key is to define the user similarity. A common way is to calculate the distance (e.g., Euclidean, Pearson correlation) between user feature vectors to measure the similarity. We utilize both the algorithm predictions and user evaluations to define a proper user similarity measure. For each user pair \((i,q)\), we can get $K$ different distances based on predictions of $K$ algorithms, denoted by \(\{d^{k}_{iq}\}\). To combine these distances together, we give each distance a confidence derived from the user performance of the base recommendation algorithms. Let \(m^{k}_{i}\) be the performance indicator of \(k\)-th algorithm on user \(i\): the smaller difference between $X^{k}_{i}$ and $Y_{i}$, the larger \(m^{k}_{i}\). Then the final distance between user \(i\) and \(q\) is defined as:
\begin{equation}
d_{Final}(i,q)= \frac{\sum_{k=1}^{K}\frac{m^{k}_{i}+m^{k}_{q}}{2}d^{k}_{iq}}{\sum_{k=1}^{K}\frac{m^{k}_{i}+m^{k}_{q}}{2}}.
\end{equation} 
The algorithms that achieve higher performances on the two users will contribute more to the final distance. Besides, the choice of performance measure \(m^{k}_{i}\) is flexible: We could select NDCG or Recall for top-N recommendation, or select MAE or RMSE for rating prediction. Note that in some metrics like MAE small values indicate high performances, we use the inverse of these metrics for \(m^{k}_{i}\). With distance function defined, we use k-medoids method to group users. 

\subsection{Ensemble Learning with Grouped Users}
For each user we learn a linear function \(f_{i}(X^{K}_{i}) = w^{T}_{i}X^{K}_{i} \) to combine the results. By now we have divided users into \(Z\) groups; users within each group share similar parameters. Then the global empirical risk function we want to minimize becomes:
\begin{equation}
\begin{split}
&\ L(W)=  \sum\limits_{i=1}^{N} l( w^{T}_{i}X^{K}_{i*}, Y_{i*}) +\Omega(W), \\
&\ \Omega(W)= \alpha \sum\limits_{z=1}^{Z}\sum\limits_{v \in \mathbb{I}_{z}} \Vert w_{v}-\bar{w}_{z} \Vert^{2} + \beta \sum\limits_{z=1}^{Z} \Vert \bar{w}_{z} \Vert^{2}, \\
\end{split}
\end{equation} 
where \(l(\cdot,\cdot)\) is the loss function, \(Y_{i*}\) indicates the ratings of user \(i\) in the training data, \(\Omega(W)\) is the regularization form and \(W = [w_{1},..., w_{N}] \) is the weight matrix to be estimated, \(\alpha\) and \(\beta\) are the regularization parameters, \(v \in \mathbb{I}_{z}\) means user \(v\) is in the group \(z\), and \(\bar{w}_{z}\) is the average weights of group \(z\). The first term in \(\Omega(W)\) tries to enforce the grouping property of \(w_{v}\). And the second term tries to avoid big values. Note that for flexibility we do not force all the users in the same group to have the same weight. However, the grouping information provides hints as to the similarities between the weights, this is clearly different from classical mixture of linear models \cite{chaganty2013spectral}, in which same groups use the same weights. 

Although eq (2) is not convex, the minimization of each group is a convex problem. We can learn the ensemble weights separately when users are already grouped.

\subsection{Incorporating New Users}
In the weighted user grouping, we use prediction metrics for each user to derive \(m^{k}_{i}\). However, new users who have few recorded ratings in \(Y\) will interfere with the grouping phase as we can not accurately measure \(m^{k}_{i}\). As a kind of cold start problem, the new user problem is a common challenge in recommendation area. 
To tackle this problem, we use an approximate approach based on a spectral relaxation form of eq (2). Specifically, the relaxation is based on regularization with spectral functions of matrices and transforms eq (2) to a convex problem. Instead of clustering users explicitly, this method implicitly clusters users by the constraints of the convex problem during learning the ensemble weights $W$.  Thus it does not need \(m^{k}_{i}\) for each user and can address the new user problem. Following the work of \cite{ding2004k}, we first reformulate the regularization \(\Omega(W)\) to:
\begin{equation}
\Omega(W,F) = \alpha(\Tr(W^{T}W) - \Tr(F^{T}W^{T}WF)) - \beta \Tr(W^{T}W),
\end{equation} 
where the matrix \(F \in \mathbb{R}^{N \times Z}\) is an orthogonal cluster indicator matrix with \(F_{i,z}=\frac{1}{\sqrt{N_{z}}}\) if \(i \in \mathbb{I}_{z}\) and \(F_{i,z}\) = 0 otherwise. Then, by ignoring the special structure of \(F\) and keeping the orthogonality requirement only \cite{zhou2011clustered}, we can transform eq(3) to:

\begin{equation}
\Omega(W,F) = \alpha \Tr(W((1+\eta) I - FF^{T})W^{T}),
\end{equation} 
where \(\eta = \beta / \alpha\). Since \(FF^{T} = I\), we rewirte the regularization as:
\begin{equation}
\Omega(W,F) = \alpha \eta (1+\eta) \Tr(W(\eta I + FF^{T})^{-1}W^{T}).
\end{equation} 
After that we can get the following convex relaxation by following \cite{chen2009convex}:
\begin{equation}
\begin{split}
&\ \min_{W,M} L(W) + \Omega_{appr}(W,M),\\
&\ ~~~~s.t. \Tr(M)=k, M \preceq I, M \in \mathbb{S}^{m}_{+,}\\
\end{split}
\end{equation} 
where \(\Omega_{appr}\) is defined as:
\begin{equation}
\Omega_{appr}(W,M) = \alpha \eta (1+\eta) \Tr(W(\eta I + M)^{-1}W^{T}),
\end{equation} 
\(\mathbb{S}^{m}_{+}\) is the subset of positive semidefinite matrices of size \(m\) by \(m\), and \(M \preceq  I\) means \(M-I\) is positive semidefinite. 

We choose an Alternating Optimization Algorithm \cite{argyriou2008convex} to solve this convex relaxation. It works by alternatively optimizing a variable with the other variables fixed. Each loop of the optimization involves the following two steps:

\textbf{Optimization of W} For a fixed M, the optimal W can be obtained via solving:
\begin{equation}
\min_{W} L(W) + c  \Tr(W(\eta I + M)^{-1}W^{T}).
\end{equation} 
We use gradient descent method to solve this convex problem. 

\textbf{Optimization of M} For a fixed W, the optimal M can be obtained via solving:
\begin{equation}
\begin{split}
&\ \min_{M} \Tr(W(\eta I + M)^{-1}W^{T}), \\
&\ ~~~~s.t. \Tr(M)=k, M \preceq I, M \in \mathbb{S}^{m}_{+.}\\
\end{split}
\end{equation} 
From \cite{zhou2011clustered}, the minimization problem equals to an eigenvalue optimization problem, the details and proofs can be found in \cite{chen2009convex}. For clarity, we call this method UREC$_{appr}$ in the latter experiments.
\subsection{Discussion}
Our ensemble problem can be regarded as a special version of semi-supervised learning: The regular users have $X^{K}$ and $Y$ available while new users only have predicted $X^{K}$. Following the idea of semi-supervised learning, the estimated mapping of regular users from $X^{K}$ and $Y$, which involves the ensemble weights $W$, can be further improved by making use of new users' predictions. However, our model is very different from tradition methods in semi-supervised learning as we focus on learning user-sensitive $W$. Currently there is no existing algorithm in semi-supervised learning to deal with our situation. As a line of future work, we will try to tailor semi-supervised learning approachs to solve our problem.

UREC can be seen as leveraging the strengths of base algorithms and discarding their weaknesses. To address the user heterogeneity problem, we prefer to combine algorithms with different structures and techniques as they are more likely to suit different types of users. This is different from traditional ensemble methods, which often focus on combining numerous but weak (sometimes homogeneous) algorithms.

\section{Experiments}
\subsection{Datasets and Settings}
We consider three public datasets for experiments: \emph{MovieLens-100K}, \emph{MovieLens-1M}, and \emph{Epinions}. They are widely experimented in the recommendation area. The details of each dataset are listed in Table 2. Besides, \emph{Epinions} has 487,145 social relations among users. To avoid data biases, we randomly select \(80\%\) of each dataset for training, \(10\%\) of each dataset for validation, and the remaining data for testing. For UREC, we equally split the validation set into two subsets. One subset is used for computing the metric \(m^{k}_{i}\), the other is used for tuning the parameters.

\begin{table}[htb]
\label{table_s}
\centering
\small
\begin{tabular}{|l|c|c|c|c|c|}
\hline
Dataset&  MovieLens-100K  & MovieLens-1M  & Epinions   \\
\hline
Users (U) & 943 & 6,040 &32,424\\ \hline
Items (V)& 1,682 & 3,900 &61,274  \\ \hline
Ratings (R)& 100,000 & 1,000,209 &664,824\\ \hline
R-Density& 6.30\% & 4.25\%& 0.03\% \\ \hline
\end{tabular}
\caption{Data statistics. \(U,V\), and \(R\) show the counts of each feature; R-Density indicates ratings links density.}
\end{table}

We then choose various prevalent methods for comparison, including: (1) KNN \cite{DBLP:conf/sigir/HerlockerKBR99}, the most common collaborative filtering algorithm that predicts users' preference based on their k-nearest neighbors.  (2) LFM \cite{bell2007modeling}, the `standard' recommendation model that utilizes matrix factorization;  (3) SVD++ \cite{koren2008factorization}, a hybrid recommendation that utilizes user implicit feedback information and ratings. It is widely applied as a benchmark; (4) TrustMF \cite{yang2016social}, as one of the state-of-the-art social recommendation models, it utilizes social information as a regularization for recommendation. We run this model on the dataset \emph{Epinions}; (5) Stacking, a traditional ensemble method that is first introduced by \cite{wolpert1992stacked}. It uses the predictions of base algorithms as input and then uses an ensemble model to predict the output. We use linear regression as the ensemble model and choose two strategies in the experiments: the first is to learn one linear regression model for all the users, denoted by Stacking$_{one}$, the second is to learn a separate linear regression model for each user, denoted by Stacking$_{user}$.

To evaluate the prediction performance, we adopt two
commonly used metrics -- mean average error (MAE) and root mean squared error (RMSE), which are defined in eq (10). The \(R_{i,j}\) denotes observed rating in testing data, \(\hat{R}_{i,j}\) is the predicted rating, and \(T\) denotes the set of tested ratings. The smaller the MAE and RMSE are, the better the rating prediction performance is.
\begin{equation}
\begin{split}
&\ MAE = \frac{\sum_{(i,j)\in T} |\hat{R}_{i,j} - R_{i,j}|}{T}, \\
&\ RMSE = \sqrt{\frac{\sum_{(i,j)\in T} (\hat{R}_{i,j} - R_{i,j})^{2}}{T}}. \\
\end{split}
\end{equation}

\subsection{Performance of Recommendation Ensemble}

\begin{table*}[thb]
\centering
\begin{tabular}{ |l|l|l|l|l|l|l|l|l| }
\hline
\multicolumn{9}{ |c| }{Effectiveness of models }  \\
\hline
Dataset & Metrics & KNN  & LFM & SVD++ & Stacking$_{one}$  & Stacking$_{user}$  & UREC$_{appr}$  & UREC \\ \hline
\multirow{2}{*}{MovieLens-100K} &  MAE& 0.7594 & 0.7329 & 0.7272 & 0.7274 & 0.7340 &  0.7257& \textbf{0.7228} \\ 
 & RMSE &  0.9649 & 0.9289 & 0.9212 & 0.9214 & 0.9322 & 0.9226 & \textbf{0.9190} \\ \hline
\multirow{2}{*}{MovieLens-1M} &  MAE& 0.7325 & 0.6869 & 0.7006 &0.6038 & 0.6027 &0.5975 & \textbf{0.5837}  \\ 
 & RMSE & 0.9221 & 0.8735 & 0.8854 & 0.7779 & 0.7706 & 0.7657 & \textbf{0.7514} \\ \hline
\multirow{2}{*}{Epinions} &  MAE& 0.8570 &  0.8399& 0.8219 & 0.8124  &0.8157 & 0.8116 & \textbf{0.8064}  \\ 
 & RMSE & 1.1466 & 1.1192 & 1.0832 & 1.0723 & 1.0705 & 1.0677& \textbf{1.0514} \\ \hline
 \end{tabular}
\caption{Performance of different models on three datasets. }
\end{table*}
We use grid search to tune the parameters to achieve the best performance. The detailed strategies are as follows: (1) For KNN, we set \(k=70\). (2) For LFM and SVD++, the learning rate is set as 0.001 and the factor dimension is set as 10. (3) For Stacking$_{one}$ and Stacking$_{user}$, we use them to combine KNN, LFM, and SVD++ for prediction. (4) For TrustMF, we set the factor dimension is set as 10 and social regularization coefficient as 0.4. (5) For UREC, we choose the inverse of RMSE to measure the \(m^{k}_{i}\). The optimal group number varies from different datasets and the details are further discussed in next subsection. For the new users, we use the average metric performance as confidence in the user grouping phase. (6) For UREC$_{appr}$, we set the \(\alpha=1\) and \(\beta=1\). The optimal group number is set the same as UREC. \emph{Epinions} has a large number of items and leads to a high time cost, and we randomly choose 5000 items to alleviate this problem.

We show the performances of our models with all the baseline models in Table 3. We can see that UREC and UREC$_{appr}$ outperform other methods on all the three datasets, according to the MAE and RMSE. KNN has the worst performances because it only utilizes user-user similarities to find neighbors and predict with a weighted average of the neighbors' ratings. LFM outperforms KNN since it integrates item-item similarities and user-user similarities by matrix factorization. By considering both user and item sides, LFM can provide more personalized predictions. SVD++ adds implicit feedback information other than ratings and can better capture the user latent factors. So it provides more accurate predictions than KNN and LFM. Stacking$_{one}$ and Stacking$_{user}$ perform better than three base models in general as they combined all these baseline models. But the difference between Stacking$_{one}$ and Stacking$_{user}$ is small. Stacking$_{one}$ is slightly better than Stacking$_{user}$, especially in \emph{Epinions}. This is very likely because the sparsity problem is severe in \emph{Epinions}. UREC$_{appr}$ is very close to UREC and also beats other models. Note that the ensemble methods (Stacking and our methods) in \emph{MovieLens-100K} did not perform much better than base algorithms compared to those in other datasets. There are two reasons: the sizes of users and items are small in \emph{MovieLens-100K}, and the rating density is high at 6.30\%. The base algorithms can learn sufficient good models with enough data, and leave less space for ensemble methods to improve. The other two datasets have a larger size and small density, so the ensemble methods gain much improvement compared to base algorithms. This verifies the necessity of recommendation ensembles and superiority of our proposed methods. 

Besides, We run TrustMF on \emph{Epinions} and use UREC to learn ensemble with TrustMF and other base models. The experimental results are shown in Table 4. TrustMF performs better than SVD++ as it utilizes the social information between users. We can see that UREC outperforms other models, indicating that TrustMF also has the `skewed prediction' problem and UREC can gain improvement on state-of-the-art algorithms.

\begin{table}[tb]
\centering
\footnotesize
\begin{tabular}{ |l|l|l|l|l|l|l|l|l| }
\hline
 \multicolumn{4}{ |c| }{Epinions - UREC  }  \\
\hline
 Methods &   SVD++ &TrustMF&  Stacking$_{user}$\\ \hline
MAE & 0.8219 & 0.7847 & 0.7903\\ \hline
 RMSE &  1.0832 & 0.8838& 0.8909 \\ \hline
 Methods & UREC & UREC$_{appr}$ &  Stacking$_{one}$\\ \hline
MAE &\textbf{0.7723} & 0.7789 &0.7803\\ \hline
 RMSE & \textbf{0.8744} & 0.8790& 0.8804 \\ \hline
\end{tabular}
\caption{Effectiveness of UREC and TrustMF on Epinions.}
\end{table}

\subsection{Analysis of User Groups and Skewness}
In this section we further discuss two issues of the user groups and skewness: (1) How does the group number \(Z\) influence the performance of UREC?  (2) Does UREC have user skewness problem? If so, how well does UREC address this problem compared to other models? 

\textbf{Analysis of User Groups}. UREC learns ensemble weights within the groups and the users in the same group are assumed to be homogeneous. So the group number \(Z\) is vital to the performance of ensemble. We experiment with different group numbers and show the corresponding results in Table 5. The performance of UREC first increases when \(Z\) gets larger. After reaching its peak at appropriate values of \(Z\), the performance decreases. The optimal \(Z\) indicates the underlying group numbers of the dataset. In \emph{MovieLens-100K}, the best \(Z\) is 3, in \emph{MovieLens-1M} the best \(Z\) is 7, and in \emph{Epinions}, the best \(Z\) is 10. Note that if all the users are homogeneous, there will be one group and our method equals to Stacking$_{one}$. In fact if we assign each user to a unique group, our method then equals to Stacking$_{user}$. From Table 5 we can see that with a proper group number our method will outperform Stacking$_{one}$ and Stacking$_{user}$. This conforms to our expectation that Stacking$_{one}$ does not consider the user heterogeneity phenomenon and it provides biased recommendation. And Stacking$_{user}$ is heavily influenced by the data sparsity problem. It will perform poorly for those users with few or no recorded interactions. Our methods utilized predictions of base algorithms to capture user heterogeneity. Other side information like user reviews also contains useful information to user heterogeneity, which is worth further exploring.

\begin{table}[tb]
\centering
\footnotesize
\begin{tabular}{ |l|l|l|l|l|l|l|l|l| }
\hline
\multicolumn{6}{ |c| }{MovieLens-100K - UREC  }  \\
\hline
 Metric & Z = 2  & Z = 3 & Z = 5 & Z = 10& Z = 20\\ \hline
MAE &  0.7295 &  \textbf{0.7228} & 07293 & 0.7301 &0.7311 \\ \hline
RMSE &  0.9250 &  \textbf{0.9197} & 0.9280 &0.9290&0.9311 \\ \hline
\hline
\multicolumn{6}{ |c| }{MovieLens-1M - UREC  }  \\
\hline
 Metric & Z = 3  & Z = 5 & Z = 7 & Z = 10& Z = 20\\ \hline
MAE &  0.6016 & 0.5924 & \textbf{0.5837} & 0.5850 & 0.6025 \\ \hline
RMSE &  0.7746 & 0.7683 & \textbf{0.7514} &0.7553 & 0.7764 \\ \hline
\hline
 \multicolumn{6}{ |c| }{Epinions - UREC  }  \\
\hline
 Metric & Z = 3  & Z = 5 & Z = 7 & Z = 10& Z = 20\\ \hline
MAE &  0.8112& 0.8104 & 0.8094 & \textbf{0.8064} &0.8232 \\ \hline
RMSE & 1.0655 & 1.0623 & 1.0573 &\textbf{1.0514} &1.0819 \\ \hline

\end{tabular}
\caption{Effectiveness of user groups \(Z\).}
\end{table}

\textbf{Analysis of User Skewness}. We have shown that UREC outperforms other models on the average based metrics. But it does not reflect the user skewness of UREC and the skewness difference compared to other models. To further explore the user skewness problem, we calculate RMSE for each user and display the statistical results in Table 6. Due to space limitations, we only show two datasets and certain some models with poor performance. We can see that in \emph{MovieLens-100K} the ensemble models have a small variance around \(0.0263\), while the variance of SVD++ is 0.0539. Stacking$_{one}$ and Stacking$_{user}$ combine several models so they are more stable than single base algorithms. When it comes to the winning rate, the UREC has a high rate around 80\%, meaning that UREC is superior to other models on most users. For traditional ensemble methods Stacking$_{one}$ and Stacking$_{user}$, they perform much worse: the wining rate of Stacking$_{one}$ is around 55\% and that of Stacking$_{user}$ is around 60\%. This finding reveals that UREC suffers little from the skewness problem compared to traditional ensemble methods. For UREC$_{appr}$, it outperforms other base models and Stacking. But the winning rate is low: it is around 55\% against Stacking$_{one}$ and Stacking$_{user}$. In \emph{MovieLens-1M}, we can observe the similar phenomenon that our methods achieve a high winning rate over other base models.
From the above analysis, we conclude that our methods can alleviate the user skewness problem and improve the recommendation performance.

\begin{table}[tb]
\centering
\scriptsize
\begin{tabular}{ |l|l|l|l|l|l|l|l|l| }
\hline
\multicolumn{6}{ |c| }{\textbf{MovieLens-100K} - UREC vs. Others }  \\
\hline
 Metric & UREC  & UREC$_{appr}$ & SVD++ & Stacking$_{one}$& Stacking$_{user}$ \\ \hline

Avg. &  0.9143 & 0.9197 & 0.9166 & 0.9142 &0.9170 \\ \hline
Var. &  0.0263 & 0.0263 & 0.0539 &0.0264&0.0272 \\ \hline
Win. & - & 78.47\% & 78.15\% & 67.32\% & 79.11\%\\ \hline
\multicolumn{6}{ |c| }{ UREC$_{appr}$ vs. Others }  \\
\hline
 Metric & KNN  & LFM & SVD++ & Stacking$_{one}$& Stacking$_{user}$ \\ \hline
Win. & 86.42\% & 79.21\% &  75.15\% & 55.33\% &53.15\% \\ \hline

\multicolumn{3}{|c|}{Stacking$_{one}$ vs. Others } & \multicolumn{3}{|c|}{Stacking$_{user}$ vs. Others}\\\hline
Metric & SVD++ & LFM &Metric & SVD++ & LFM\\ \hline
Win. & 55.99\% &62.30\% & Win. &57.51\%& 61.22\%\\ \hline
\hline
\multicolumn{6}{ |c| }{\textbf{MovieLens-1M} - UREC vs. Others }  \\
\hline
 Metric & UREC  & UREC$_{appr}$ & SVD++ & Stacking$_{one}$& Stacking$_{user}$ \\ \hline

Avg. &  0.6706 & 0.6889 & 0.8288 & 0.6973 &0.6977 \\ \hline
Var. &  0.0482 & 0.0501 & 0.2684 &0.0503 &0.0528 \\ \hline
Win. & - & 67.29\% &  58.81\%&69.22\% & 68.97\% \\ \hline

\multicolumn{6}{ |c| }{ UREC$_{appr}$ vs. Others }  \\
\hline
 Metric & KNN  & LFM & SVD++ & Stacking$_{one}$& Stacking$_{user}$ \\ \hline
Win. & 65.47\% & 55.45\% &  56.84\% & 51.78\% &51.35\% \\ \hline

\multicolumn{3}{|c|}{Stacking$_{one}$ vs. Others } & \multicolumn{3}{|c|}{Stacking$_{user}$ vs. Others}\\\hline
Metric & SVD++ & LFM &Metric & SVD++ & LFM\\ \hline
Win. & 54.22\% &50.81\% & Win. &55.94\%& 54.93\%\\ \hline

  \end{tabular}
\caption{Performance of RMSE on user level. Avg. means the average RMSE of users, Var. means the variance of RMSE of users, and Win. (winning rate) means the winning percentage of users that the stated algorithm versus other base models.}
\end{table}

\subsection{Comparison of UREC and UREC$_{appr}$}
We next compare our proposed UREC and UREC$_{appr}$ in two aspects: accuracy and scalability. (1) \textbf{Accuracy}. From the experimental analysis in Sections 4.2 and 4.3, UREC achieves better accuracy than UREC$_{appr}$ in general. For the user skewness problem, UREC also gets a higher winning rate. Note that UREC$_{appr}$ also outperforms other base models, and its difference from UREC is minor. UREC$_{appr}$ can be an alternative method.
 (2) \textbf{Scalability}. When datasets become larger, the high complexity of the clustering procedure will be a bottleneck of our ensemble methods. UREC is slow in efficiency as k-medoids method has a high complexity. UREC$_{appr}$ transforms the problem to a convex relaxed problem. Table 7 shows the elapsed time for training UREC and UREC$_{appr}$ , and UREC$_{appr}$ is indeed faster. Note that the runtime for both methods increases dramatically when the datasets become larger. Recent studies have proposed several efficient optimization methods to address this problem \cite{zhou2011clustered}. Moreover, UREC$_{appr}$ can handle the new user problem that is a prevailing issue in online websites. As a consequence, one can say that the UREC$_{appr}$ is more scalable.

\begin{table}[tph]
\centering
\scriptsize

\begin{tabular}{ |c|c|c|c|c|}
\hline
\multicolumn{4}{ |c| }{Runtime Comparison (In Seconds)  }  \\
\hline
Method  & MovieLens-100K& MovieLens-1M & Epinions\\ \hline
UREC  & 452 \(\pm\) 4.0 & 6152 \(\pm\) 8.5  & 41023\(\pm\) 36.0 \\ \hline
UREC$_{appr}$  & 265 \(\pm\) 2.5& 4323 \(\pm\) 6.0  & 25404\(\pm\) 23.0 \\
\hline
\end{tabular}
\caption{Runtime comparison of UREC and UREC$_{appr}$. }
\end{table}

\section{Conclusion}
In this paper we proposed a novel method for user-sensitive recommendation ensemble called UREC to address the user skewed prediction problem. The proposed method has a clear intuition. UREC first clusters users based on the predictions of base recommendation algorithms, and then it uses multi-task learning to learn the ensemble weights. To alleviate the new user problem that usually interferes with user grouping, we propose an approximate approach named UREC$_{appr}$ based on a spectral relaxation of regularization. Empirical results on three benchmark real-world datasets show that our methods clearly outperform alternatives. Further experimental results demonstrate UREC can better alleviate the user skewness problem than traditional ensemble methods and improve the recommendation performance. In addition, UREC$_{appr}$ also achieves a competitive and promising performance. Those empirical results verifies the necessity of developing use-sensitive ensembles and the efficacy of the proposed ensemble scheme. We believe that the proposed method and the observations in experimental results will inspire more approaches to recommendation ensembles.
In the future, we will investigate how to leverage side information (e.g., social relations and user annotated tags) to better capture user heterogeneity.
\newpage

\bibliographystyle{named}
\bibliography{ijcai18}

\end{document}